\documentclass[table,acmsmall,screen]{acmart}
\usepackage{subcaption}
\usepackage{multirow} 
\usepackage{booktabs}
\usepackage{caption}
\usepackage{textcomp}
\usepackage{multirow}
\usepackage{caption}
\usepackage{xcolor}
\usepackage[ruled,linesnumbered]{algorithm2e}
\usepackage{enumitem}
\usepackage[most]{tcolorbox}
\usepackage{balance}
\usepackage{url}
\usepackage{pifont} %
\usepackage{tabularx}
\usepackage{graphicx}
\usepackage{array}
\usepackage{longtable}
\usepackage{multirow}
\usepackage{caption}
\usepackage{xcolor}
\usepackage[ruled,linesnumbered]{algorithm2e}
\usepackage{enumitem}
\usepackage{balance}
\usepackage{url}
\usepackage{booktabs}
\usepackage{framed}
\usepackage{cellspace}
\usepackage{colortbl}
\usepackage[most]{tcolorbox}

\usepackage{pifont}
\usepackage[ruled,linesnumbered]{algorithm2e}
\usepackage{algorithmic}
\algsetup{linenosize=\small}
\SetKwInput{KwInput}{Input}                %
\SetKwInput{KwOutput}{Output}              %

\usepackage[ruled,linesnumbered]{algorithm2e}
\usepackage{algorithmic}
\algsetup{linenosize=\small}
\SetKwInput{KwInput}{Input}                %
\SetKwInput{KwOutput}{Output}              %
\usepackage{quoting}

\usepackage{xcolor}
\usepackage{enumitem}

\definecolor{mygrey}{HTML}{d3d3d3}
\definecolor{myblue}{HTML}{F5DDFE}
\definecolor{mygreen}{HTML}{CEF5D3}
\newtcolorbox{boxA}{
    boxrule = 1.5pt,
    colback = myblue, %
    boxsep=0.1pt,
    top=3pt,bottom=2pt,
    boxrule=1pt,
    frame hidden,
    sharp corners,
    enhanced,
    borderline north={1pt}{0pt}{black},
    borderline south={1pt}{0pt}{black},
    boxsep=2pt,left=2pt,right=2pt,top=2.5pt,bottom=2pt
}

\newtcolorbox{boxB}{
    boxrule = 1.5pt,
    rounded corners,
    colback = mygreen, %
    boxsep=0.1pt,
    top=3pt,bottom=2pt,
}

\AtBeginDocument{%
  \providecommand\BibTeX{{%
    \normalfont B\kern-0.5em{\scshape i\kern-0.25em b}\kern-0.8em\TeX}}}

\setcopyright{acmcopyright}
\acmJournal{CSUR}
\acmYear{2023} \acmVolume{1} \acmNumber{1} \acmArticle{1} \acmMonth{10} \acmPrice{15.00}\acmDOI{xxx }

\acmJournal{JACM}
\acmVolume{37}
\acmNumber{4}
\acmMonth{10}

\begin{document}

\title{Ecosystem of Large Language Models for Code}

\author{Zhou Yang}
\email{zyang@smu.edu.sg}
\affiliation{%
  \institution{Singapore Management University}       
  \country{Singapore}
}

\author{Jieke Shi}
\email{jiekeshi@smu.edu.sg}
\affiliation{%
  \institution{Singapore Management University}
  \country{Singapore}
}

\author{Premkumar Devanbu}
\email{ptdevanbu@ucdavis.edu}
\affiliation{%
  \institution{Department of Computer Science, UC Davis}
  \country{USA}
}

\author{David Lo}
\email{davidlo@smu.edu.sg}
\affiliation{%
  \institution{Singapore Management University}
  \country{Singapore}
}
\renewcommand{\shortauthors}{Yang et al.}

\begin{abstract}
  The extensive availability of publicly accessible source code and the advances in language models, coupled with increasing computational resources, have led to a remarkable rise of large language models for code (LLM4Code).
  These models do not exist in isolation but rather depend on and interact with each other, forming a complex ecosystem that is worth studying.
  It motivates us to introduce a pioneering analysis of the \textit{LLM4Code ecosystem}. 
  Utilizing Hugging Face \includegraphics[height=1em]{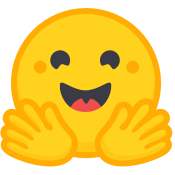}—the premier hub for transformer-based models—as our primary source, we manually curate a list of datasets and models focused on software engineering tasks.
  We first identify key datasets, models, and users in the ecosystem, and quantify their contribution and importance.
  We then examine each model's documentation to trace its base model and understand the process for deriving new models.
  We categorize LLM4Code model reuse into nine categories, with the top three being \textit{fine-tuning}, \textit{architecture sharing}, and \textit{quantization}.
  Additionally, we examine documentation and licensing practices, revealing that LLM4Code documentation is less detailed than that of general AI repositories on GitHub.
  The license usage pattern is also different from other software repositories, and we further analyze potential license incompatibility issues.
  To analyze the rapidly growing LLM4Code, we explore the potential of using LLMs to assist in constructing and analyzing the ecosystem.
  Advanced LLMs from OpenAI identify LLM4Code with 98\% accuracy, infer base models with 87\% accuracy, and predict reuse types with 89\% accuracy.
  We employ LLMs to expand the ecosystem and find conclusions from the manually curated dataset align with those on the automatically created one.
  Based on our findings, we discuss the implications and suggestions to facilitate the healthy growth of LLM4Code.
\end{abstract}


\maketitle

\section{Introduction} \label{sec:introduction}

The development of \textit{LLM4Code}~\cite{CodeBERT,codegen,wang2021codet5} has benefited from the easy access to extensive public source code datasets~\cite{husain2019codesearchnet,CodeXGLUE} and significant progress in modern language models~\cite{bert,RoBERTa,gpt-2}, along with growing computational resources.
Such rapid development also owes a great deal to the open-source culture.
Many models are made available to the public (e.g., shared on model zoos like Hugging Face), which are reused and extended to derive new models.

LLM4Code do not exist in isolation but rather depend on and interact with each other.
For example, the model \path{intm/codet5-small-go_generation}\footnote{\url{https://huggingface.co/intm/codet5-small-go\_generation}} is derived from the base model \path{Salesforce/codet5-small}, which, in turn, shares the same architecture with \path{t5-small}.
This is similar to other software ecosystems. 
For instance, in the PyPi ecosystem~\cite{9284011,9425974,9284138}, one package may depend on another package.
Researchers have analyzed the dependencies of various ecosystems to understand software evolution~\cite{9794026}, security~\cite{9284011,massacci2021technical,10123571}, and licensing issues~\cite{10.1007/978-3-031-08129-3_6}.
This motivates us to offer the first exploration of LLM4Code from an ecosystem perspective. 

We define the \textit{LLM4Code ecosystem} as a dynamic network composed of multiple LLMs and datasets focused on software engineering tasks. 
Within this ecosystem, models interact through complementary and derivative relationships. 
These models often originate from foundational models and evolve into new models tailored for specific software engineering tasks through various model reuse including fine-tuning, quantization, and task-specific enhancement strategies like adversarial training. 
Each model also contributes data and knowledge to support other models, creating a mutually dependent and reinforcing supply chain. 
Over time, these models continuously influence and improve each other, driving the entire ecosystem toward greater efficiency, intelligence, and adaptability, ultimately providing robust support for tasks such as code generation, program repair, software testing, and documentation generation.

To build the LLM4Code ecosystem, we use keyword search to fetch a list of models and datasets hosted on Hugging Face, the most prominent model hub~\cite{jiang2023empirical,ptmtorrent}. 
We manually review their documentation to include only those designed for software engineering tasks, which we refer to as \textit{entities} in the ecosystem.
Subsequently, we manually document the dependencies among these entities.
This process yields an ecosystem consisting of 366 models and 73 datasets, with 495 dependencies among them.
The value of this ecosystem is demonstrated through the exploration of three research questions (RQs).

We first identify the important datasets, models, and contributors in the ecosystem (RQ1).
Users on Hugging Face can like models if they find them useful. The numbers of `like' are recorded, which can measure models' popularity. 
We identify key contributors by summing the total likes on all models they own.
We find that the top-3 contributors are \path{Salesforce}, \path{bigcode}, and \path{microsoft}.
We perform statistical tests to highlight the prominence of company accounts (i.e., the accounts marked as company or enterprise) and company-owned models in the ecosystem. 
Measured by the average number of likes, company-owned models are 60.2 times as popular as non-company-owned models.
We then analyze the model importance, which we measure using the number of derivatives produced from a model.
We only find a weak positive correlation between the number of dependents and likes ($p<0.01$), suggesting that a highly-reused model does not necessarily receive a proportionate amount of recognition from the community, and vice versa.
This indicates a potential disconnect between perceived value and actual impact of LLM4Code.

Then, we investigate the practices (by the larger developer community)  of reusing and deriving LLM4Code in RQ2.
We conduct open card sorting to build an LLM4Code reuse taxonomy consisting of nine distinct categories.
In descending order of prevalence, the categories are: fine-tuning (205 instances), architecture sharing~(84), quantization~(32), continue training~(15), model conversion~(14), distillation (12), Adapter~(9), instruction tuning~(9), and adversarial tuning~(6). 
The remaining 49 dependencies are considered as unclear.
Fine-tuning and architecture sharing are unsurprisingly common.
However, the rise of quantization and distillation points to a growing demand for deploying LLM4Code on resource-limited devices.
Model conversion arises from the users' need for cross-platform model operability.
Both adapter~\cite{hu2021lora} and instruction tuning represent parameter-efficient tuning strategies~\cite{weyssow2024exploring} to adapt models to new tasks and datasets with less computational cost.
Lastly, adversarial tuning emerges as a contemporary approach focused on enhancing model robustness~\cite{yang2024robustness}.
This analysis provides LLM4Code researchers and practitioners with practical insights in applying suitable strategies to produce new models.

Third, RQ3 analyzes two important aspects of practices when publishing LLM4Code:~(1) documentation and (2) license usage.
We find that the LLM4Code documentation is generally not well maintained and contains less information than that of AI repositories hosted on GitHub (compared with a collection of 1,149 AI-oriented repositories collected by Fan et al.~\cite{fan2021makes}).
However, documentation quality is a key factor to the success of OSS projects~\cite{10.1007/s10664-018-9660-3}, calling for more attention to enhance the documentation quality of LLM4Code.
Additionally, we find that over 60\% models (220 out of 366) provide no license information.
The license distribution is different compared to traditional software ecosystems.
On the one hand, LLM4Code developers tend to choose permissive licenses, e.g., MIT, Apache, and CC-BY.
On the other hand, some LLM4Code adopt AI-specific licenses, e.g., RAIL (Responsible AI Licenses)\footnote{\url{https://huggingface.co/blog/open_rail}} and AI model license agreement by AI2.\footnote{\url{https://allenai.org/impact-license}}
Using the Linux Foundation's OSS license compatibility table~\cite{licensetable}, we examine the licenses of model pairs with dependencies and found no cases of incompatibility.
However, this does not eliminate the risk of license conflicts. The absence of incompatibility is largely due to missing license information, and the compatibility table is not well-suited for AI-specific licenses.
For instance, some widely-used models, like \path{Salesforce/codet5-base}, do not provide license details. 
Model licenses also affect the dependent models. 
If upstream models update their licenses, it could impact many others in the ecosystem. 
Updating licenses is common in other software ecosystems~\cite{10.1007/s10664-024-10486-0,10.1007/978-3-031-08129-3_6}.

As the ecosystem expands rapidly, manual curation and analysis is rather expensive.
Thus, we use LLMs to assist in building and analyzing the ecosystem. 
We find that LLMs by OpenAI can identify LLM4Code with 98\% accuracy, infer the base model with 87\% accuracy, and predict reuse type with 89\% accuracy. 
Although it still requires human intervention, the results show the potential of LLMs in automating the process. 
We employ LLMs to expand the ecosystem to include 6,051 LLM4Code and empirically show that the conclusions from the manually curated dataset can generalize to the automatically created one.

The main contributions of this paper include:
\begin{enumerate}[leftmargin=*]
    \item \textbf{Dataset:} We manually curate a dataset of the code model ecosystem on Hugging Face, including 366 LLM4Code, 72 code datasets, and 495 dependencies.
    \item \textbf{Analysis:} We analyze the trends and important contributors, models, and datasets in the ecosystem.
    \item \textbf{Practices:} We create a taxonomy to understand the practices of model reuse in the ecosystem. We also analyze the practices followed when publishing LLM4Code, including documentation and license usage.
    \item \textbf{Automation:} We show the potential of using LLMs to assist in constructing and analyzing the ecosystem at scale, expanding the model size by 16.5 times.
  \end{enumerate}

  \vspace*{0.1cm}
\noindent \textbf{Paper Structure.} 
The paper provids the background in Section~\ref{sec:background}.
Section~\ref{sec:data} details the manual data collection and labeling methods.
Section~\ref{sec:research_questions} presents the research questions that showcase the values of the ecosystem.
Section~\ref{sec:automation} explains how to use LLMs to assist in constructing and analyzing the ecosystem.
In Section~\ref{sec:discussion}, we explore potential implications and discuss the threats to validity.
Section~\ref{sec:related} presents notable findings in the field.
Finally, our paper concludes with Section~\ref{sec:conclusion}, where we also outline future research directions.

\section{Background} \label{sec:background}

\subsection{Datasets and Language Models of Code}
Over the past few years, as large language models (LLMs) have achieved phenomenal success in natural language processing, the foundational architecture called \textit{transformer}~\cite{vaswani2017attention} has also been applied to process source code, producing powerful \textit{LLM4Code} like CodeBERT~\cite{CodeBERT} and StarCoder~\cite{starcoder}.
Empirical studies~\cite{niu2023empirical,10.1145/3533767.3534390,he2023representation} have demonstrated that these transformer-based LLM4Code achieve state-of-the-art performance in various software engineering tasks, such as code generation, code summarization, etc.

The success of LLM4Code largely stems from the extensive code datasets used for training and evaluation LLM4Code across different tasks. 
CodeXGLUE~\cite{CodeXGLUE}, for example, offers a comprehensive benchmark for code comprehension and generation tasks. 
Recently, code datasets of considerable magnitude, ranging from hundreds of gigabytes to even terabytes in size, have been made available. 
One notable example is The Stack,\footnote{\url{https://huggingface.co/datasets/bigcode/the-stack}} which houses an impressive 6 terabytes of source code files, covering a vast array of 358 programming languages.
These datasets produce many advanced LLM4Code like Starcoder~\cite{starcoder,starcoder2}, CodeLlama, CodeGen~\cite{codegen}, etc.
We refer interested readers to a recent survey~\cite{codellm_survey} for a comprehensive overview of code datasets and models.

\begin{figure}[!t]
    \centering
    \includegraphics[width=0.8\textwidth]{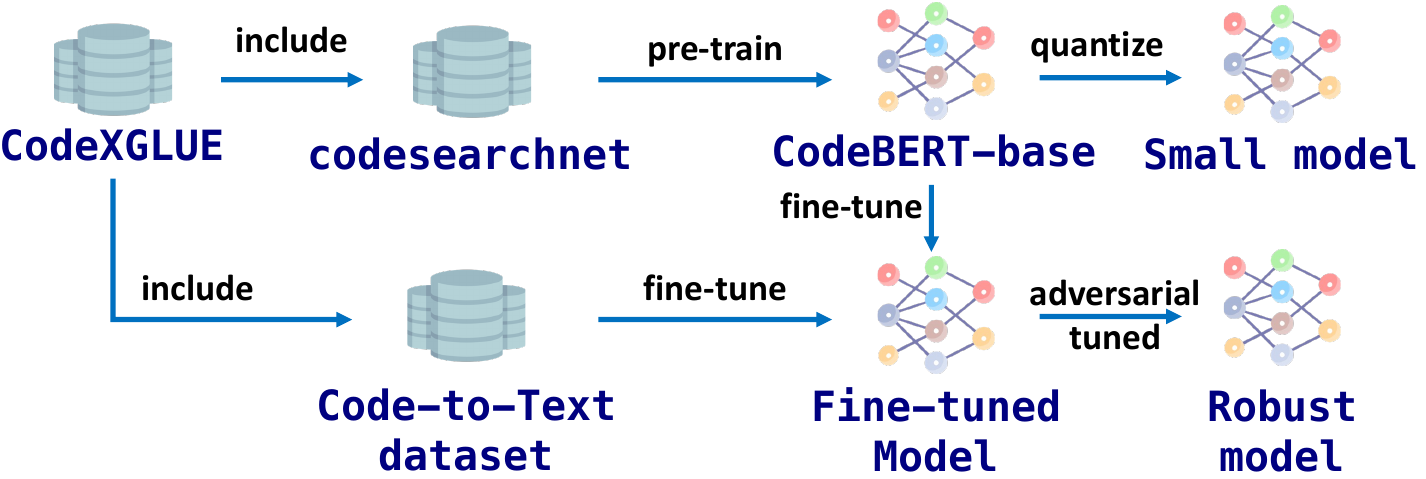}
    \caption{An illustrative example of the code model ecosystem, including code datasets, models and dependencies.}
    \label{fig:overview}
  \end{figure}

\subsection{Model Hubs}

Both researchers and engineers use \textit{model hubs} to share and reuse AI datasets and models.
Training state-of-the-art models is costly, requiring significant computational resources and time.
For instance, training GPT-3~\cite{gpt-3} on 1,024 A100 GPUs can take around a month.
Thus, a common practice is to reuse existing models, e.g., by fine-tuning it on a specific task with a small amount of data, which is more computationally friendly and still delivers promising results.
Model hubs have become essential resources for sharing and reusing pre-trained models and datasets.

Jiang et al.~\cite{jiang2022empirical} analyze eight model hubs, including: (1) Hugging Face, (2) TensorFlow Hub, (3) Model Zoo, (4) Pytorch Hub, (5) ONNX Model Zoo, (6) Modelhub, (7) NVIDIA NGC, and (8) MATLAB Model Hub.
Hugging Face is the only open model hub that allows contribution from any registered users.
Besides, with a staggering user base exceeding 1 million individuals, Hugging Face has undeniably established itself as the go-to choice for developers and researchers alike. 
Their \texttt{transformers} library (used to access and reuse models) sees an average of 300,000 daily pip installs, showing the high demand for their tools and platforms.
Thus, this paper focuses on the code datasets and LLM4Code hosted on Hugging Face.

\subsection{Software Ecosystems}

\textit{Software ecosystem}, introduced by Messerschmitt and Szyperski~\cite{messerschmitt2003software} in 2003, has inspired a large body of research.
While a singular definition remains elusive, the core idea is that software should be viewed within the larger, intertwined networks they form, characterized by dependencies and co-evolution of software components and systems.
Researchers have studied various aspects of different software ecosystems.
Common software ecosystems include PyPI~\cite{9284011,9425974,9284138}, npm~\cite{8453117,9794026}, Maven~\cite{10123571,10123641,massacci2021technical}, Doceker~\cite{7962382}, etc.
Analyzing the ecosystems leads to insights about software vulnerabilities, bugs, technical debts, license, etc. 

The concept of software ecosystem can extend to AI models.
Bommasani et al.~\cite{fundation-models} propose the concept of \textit{foundation models}, which are trained on broad data at scale and produce to a wide range of downstream models.
Bommasani et al.~\cite{ecosystem-graph} then propose the `ecosystem graph' to analyze the social footprint of foundation models. 
Similarly, code datasets and LLM4Code form an interconnected ecosystem.
Figure~\ref{fig:overview} illustrates an example of their interactions.
\texttt{CodeXGLUE} includes two datasets: \texttt{codesearchnet} and \texttt{code2nl}.
The former is used to pre-train \texttt{CodeBERT-base}, which is then used to be fine-tuned on \texttt{code2nl} to obtain a new model that is capable of summarizing source code into natural language.
This fine-tuned model is then reused to create a quantized version for resource-constrained devices and an adversarially tuned version, robust to attacks.
These interactions and dependencies motivate the first exploration of the \textit{LLM4Code ecosystem}, briefly defined as follows.
\vspace*{0.2cm}
\begin{quote}
\hangindent=0.5cm
\hangafter=1
\textit{"An LLM4Code ecosystem is a dynamic network composed of multiple LLMs and datasets focused on software engineering tasks."}
\end{quote}
\vspace*{0.2cm}
Within this ecosystem, entities interact through complementary and derivative relationships. 
Foundational models evolve into new models via reuse strategies like fine-tuning, quantization, and adversarial training.
Each model is designed and trained for its designated tasks while also contributing data, knowledge, or strategies to support other models, creating a mutually dependent and reinforcing model supply chain. 
Over time, these models continuously influence and improve each other, driving the ecosystem toward greater efficiency, intelligence, and adaptability, supporting more software engineering tasks.
This paper builds and analyzes the LLM4Code ecosystem, providing insights into the trends, practices, and advice to improve quality of the ecosystem.

\section{Manual Data Collection}
\label{sec:data}

We employ two strategies to construct and analyze the LLM4Code ecosystem: (1) manual analysis and (2) LLM-assisted automated analysis.
The advantage of manual analysis is that it allows us to precisely analyze the data (e.g., identifying reuse types) in detail.
However, it is rather expensive to adapt to the rapidly growing LLM4Code.
In this section, we first describe the manual construction process, followed by the automated dataset expansion in Section~\ref{sec:automation}.
In a nutshell, our manual analysis to construct the ecosystem can be divided into three steps:

\begin{enumerate}[leftmargin=*]
  \item We initiate the process by querying the Hugging Face API, enabling us to acquire an initial list of code datasets and models. 
  \item Next, we employ the information in the model documentation to keep only relevant datasets and models. Inspired by snowballing adopted in literature review, we systematically discover additional datasets and models.
  \item Additionally, we manually confirm the dependencies among various datasets and models. 
\end{enumerate}

\subsection{Finding Code Datasets and Models}
\label{subsec:step1}

\begin{figure*}[!t]
  \centering
  \includegraphics[width=1\textwidth]{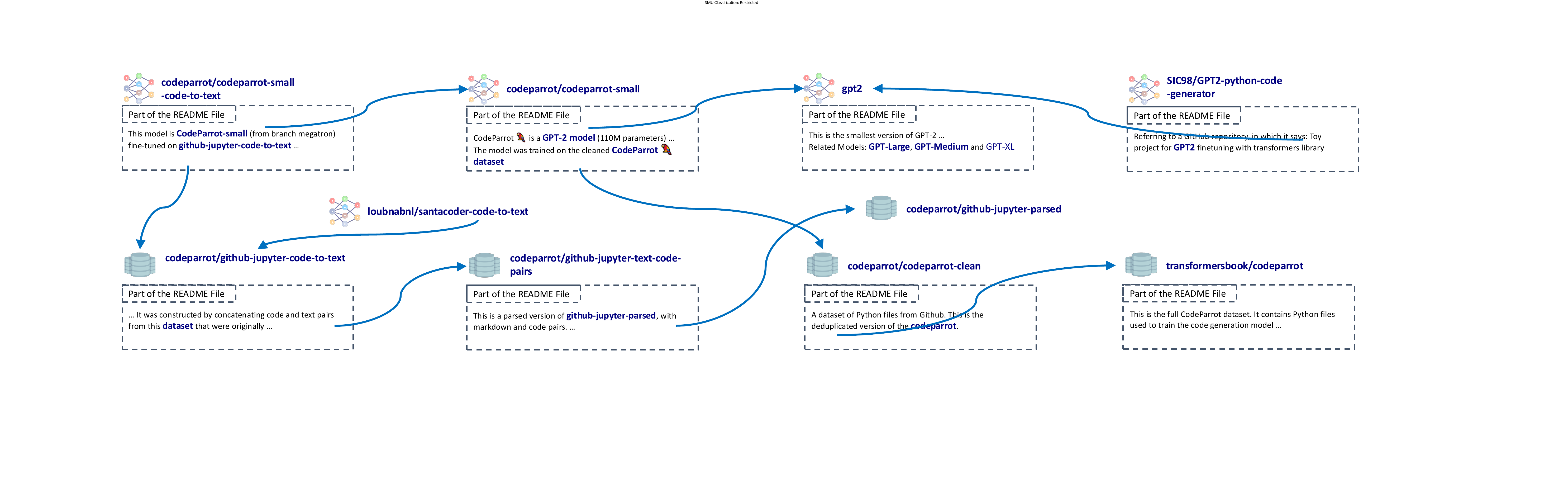}
  \caption{Examples of the dependencies between code datasets and models. The entities with the icon~\includegraphics[height=1em]{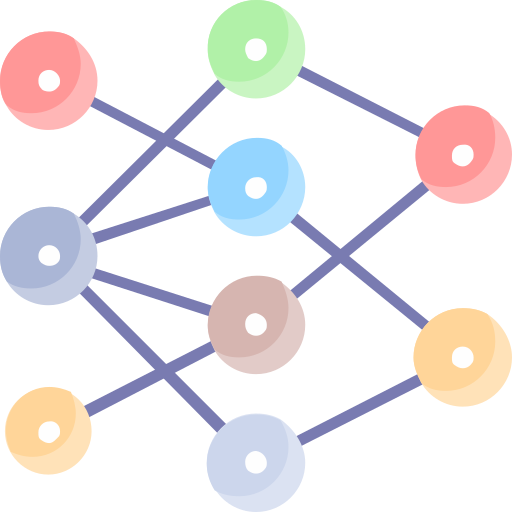} and \includegraphics[height=1em]{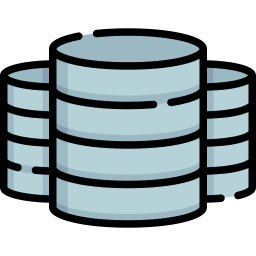} are models and datasets, respectively. The arrows indicate the dependencies between them, which are extracted from the documentation and model cards of each entities. For example, the model \protect\path{codeparrot/codeparrot-small-code-to-text} explicitly states that it fine-tunes the model \protect\path{codeparrot/codeparrot-small} on the dataset \protect\path{codeparrot/github-jupyter-code-to-text}.}
  \label{fig:example}
\end{figure*}

We utilize the Hugging Face API\footnote{\url{https://Hugging ßFace.co/docs/hub/api}} to obtain an initial list of models that are potentially relevant to software engineering.
Specifically, given our focus on code and considering that the naming conventions in this domain to include the keyword `\textbf{code}' (e.g., \textbf{Code}BERT~\cite{CodeBERT} and Santa\textbf{Code}r~\cite{santacoder}), we conduct a search using `\textbf{code}' as the keyword.
This search yields a collection of 1,566 models as of $13^{th}$ August, 2023, whose model name or owner name contains the substring `\textbf{code}.'
To ensure the models are truly code-related, we manually review the documentation for each one.

Models on Hugging Face are hosted as Git repositories, which automatically display a documentation page if a README file is present.
The documentation page, also known as the \textit{model card},\footnote{\url{https://Hugging Face.co/docs/hub/model-cards}} contains two key elements: (1) metadata in YAML and (2) descriptions in Markdown.
We carefully examined the documentation (if available) to exclude any erroneously included models.
This process refines the model list, ensuring only code-related models are included, excluding models like \path{nateraw/codecarbon-text-classification} or \path{AkshatSurolia/ICD-10-Code-Prediction}, which despite containing the keyword, focus on sentiment analysis in movie reviews and clinical documents, respectively.
In this step, we manually categorize models into three groups: (1) those with no or insufficient documentation, (2) those with documentation showing they are not code-related, and (3) those with documentation confirming they are relevant to code. Only models in the third group are retained.

Two authors independently labeled the data, spending a total of 11 and 13 hours, respectively. 
Their inter-rater agreement was calculated to be 0.77, indicating substantial agreement~\cite{mchugh2012interrater}.
In cases of disagreement, an additional annotator with LLM4Code expertise was consulted to resolve discrepancies and make the final decision. Of the 1,566 models, 320 were confirmed as LLM4Code, 85 as non-code-related models, and 1,161 lacked sufficient documentation to determine their type.

\subsection{Snowballing}
\label{subsec:snowballing}

In the previous steps, we excluded models with names containing the keyword `code' that are not LLM4Code.
But we notice that some LLM4Code do not contain the keyword in their, names. 
For example, \path{emre/java-RoBERTa-Tara-small} is omitted in the previous step but is trained on the code dataset \path{code_search_net}, indicating that it is an LLM4Code.

Since keyword searching may overlook some LLM4Code, we use \textit{snowballing} to identify additional models and ensure comprehensive LLM4Code coverage. 
Snowballing~\cite{snowball1,snowball2}, commonly used in systematic literature reviews, involves both \textit{backward} and \textit{forward} search procedures.
We tailor the snowballing method to expand our ecosystem.
For each known LLM4Code, we manually review its documentation to identify the training dataset and related models, such as those used for fine-tuning.
Hugging Face provides the datasets used to train a model and the models that use a dataset, if such information is specific in the metadata.
We process the metadata to automatically find a model's training datasets.
After confirming that a dataset is code-related, we add it to our dataset list and extracts models that use this dataset.
We further read the documentation of these newly found models to identify LLM4Code.
The process is repeated until no new LLM4Code or datasets are found.
This step yields 46 new LLM4Code.

\subsection{Manual Labeling of Dependencies}
\label{subsec:step2}

We review the documentation of each confirmed LLM4Code and dataset to understand their relationships with others.
The documentation may include external links, such as URLs to GitHub repositories or research papers. 
We also review these links to identify dependencies when necessary.
Figure~\ref{fig:example} shows a small set of datasets and models in our ecosystem, as well as the dependencies among them.
In the documentation of \path{codeparrot/codeparrot-small}, it clearly states that `\textit{CodeParrot is a GPT-2 model (110M parameters) trained to generate Python code.}'
It indicates that this model shares the GPT-2 architecture, which creates a dependency between the \path{codeparrot/codeparrot-small} and \path{gpt2}.\footnote{The GPT-2 model is called \texttt{gpt2} (\url{https://Hugging Face.co/gpt2}) on Hugging Face.}
It also mentions that `\textit{The model was trained on the cleaned CodeParrot (an URL) dataset ...},' indicating a dependent relationship between the model \path{codeparrot/codeparrot-small} and its training dataset \path{codeparrot/codeparrot-clean}.

Another model,
explicitly mentions `\textit{this model is codeparrot-small fine-tuned on github-jupyter-code-to-text.}'
From this description, we can infer a model-to-model dependency between \path{codeparrot/codeparrot-small} and the subject model, as well as a model-to-data dependency between the subject model and the dataset \path{codeparrot/github-jupyter-code-to-text}.
This dataset's documentation writes `\textit{It was constructed ... from this dataset codeparrot/github-jupyter-text-code-pairs} ...,' indicating a link to another dataset, which says `\textit{This is a parsed version of github-jupyter-parsed ...}.'
This enables us to build dataset-dataset dependencies.

After carefully reading the documentation and the external URLs they refer to (e.g., GitHub repositories or publications), we obtain 435 model-to-model dependencies and 60 model-to-data dependencies.
We then conduct open-card sorting to categorize the dependencies (Section~\ref{subsec:rq2}), which takes a total of 7 hours.

\section{Research Questions and Results}
\label{sec:research_questions}
This section showcases the value of constructing and analyzing the LLM4Code ecosystem by answering the following three research questions (RQs):
\begin{itemize}[leftmargin=*]
    \item \textbf{RQ1}. \textit{What are the popular and important datasets, models, and contributors in the ecosystem?}
    \item \textbf{RQ2}. \textit{What are the practices of reusing LLM4Code?}
    \item \textbf{RQ3}. \textit{What are the practices of publishing LLM4Code?}
\end{itemize}

\begin{table*}[!t]
  \centering
  \caption{Statistics of models and owners in the LLM4Code ecosystem.}
  \label{tab:my_table}
  {
  \begin{tabular}{ccccc}
    \toprule
    \multirow{2}{*}{\textbf{Rank}} & \multicolumn{2}{c}{\textbf{O-Network}} & \multicolumn{2}{c}{\textbf{M-Network}} \\
    \cmidrule(lr){2-3} \cmidrule(lr){4-5}
    & \textbf{Name} & \textbf{Degree} & \textbf{Name} & \textbf{Degree} \\
    \midrule
    \rowcolor{gray!15} 1 & SEBIS & 146 & t5-small & 33 \\
    2 & Salesforce & 33 & bigcode/santacoder & 30 \\
    \rowcolor{gray!15} 3 & mrm8488 & 21 & t5-base & 28 \\
    4 & codeparrot & 19 & Salesforce/codet5-base & 21 \\
    \rowcolor{gray!15} 5 & Nan-Do & 16 & bigcode/starcoder & 18 \\
    \bottomrule
    \toprule
    \multirow{2}{*}{\textbf{Rank}} & \multicolumn{2}{c}{\textbf{No. Likes}} & \multicolumn{2}{c}{\textbf{No. Total Likes for An Account}} \\
    \cmidrule(lr){2-3} \cmidrule(lr){4-5}
    & \textbf{Name} & \textbf{Likes} & \textbf{Name} & \textbf{Number} \\
    \midrule
    \rowcolor{gray!15} 1 & bigcode/starcoder & 2,362 & bigcode & 3,583 \\
    2  & replit/replit-code-v1-3b & 678 & Salesforce & 599 \\
    \rowcolor{gray!15} 3 & bigcode/santacoder & 298 & replit &  533 \\
    4 & microsoft/codebert-base & 133 & microsoft & 138 \\
    \rowcolor{gray!15} 5 & Salesforce/codegen-16B-mono & 115 & facebook & 155 \\
    \bottomrule
  \end{tabular}
  } %
\end{table*}

\subsection{RQ1. What are the popular and important datasets, models, and contributors in the ecosystem?}

\subsubsection{Motivation}

Understanding the popularity and importance of different entities in the ecosystem is crucial. 
First, it can help users make informed decisions.
Imagine new users who want to start using or contributing LLM4Code.
An important question of their interest can be: \textit{where should I start?}
They can choose to follow the community's choice, starting with the frequently liked and reused ones.
Second, popularity and importance quantify the contributions of different users, facilitating the understanding of LLM4Code development.

Besides, it helps us identify potential risks and issues in the ecosystem.
The models and dataset do not exist in isolation but are deeply interconnected through complex dependencies and interactions. 
A model may produce many downstream models, showing its importance. 
However, it also means that changes in the model may affect many downstream models. 
For example, if such an important model changes its license (which is found common in OSS projects~\cite{10.1007/978-3-031-08129-3_6}) to a more restrictive one, it may cause a chain reaction and affect many downstream models.
Or if some datasets are found poisoned~\cite{advdoor,Sun2023backdoor}, it may affect many models that are trained on them.
Understanding dependencies is crucial for understanding the stability and evolution of the ecosystem.

\subsubsection{Methodology} 
We answer this research question by conducting the following two analyses: (1) Hugging Face popularity metric analysis, and (2) dependency network analysis.
The former quantifies the popularity and the latter quantifies the importance of datasets, models, and contributors.

\vspace*{0.2cm}
\noindent \textbf{Hugging Face Popularity Metric Analysis.}
On Hugging Face, users can like and download models or datasets they find useful or want to develop further, similar to GitHub's 'star' and 'fork' features. 
The numbers of `like' and `download' are recorded, which serves as indicators of a model's popularity and user satisfaction. 
A higher number of likes suggests that the model or dataset is well-regarded by the community, reflecting its utility, performance, or relevance to users' needs.
A higher download count (often indicative of collaborative or derivative development efforts) indicates that the resource is widely adopted and applied across various projects and research areas.

Note that Hugging Face only provides the `number of model downloads \textbf{in the last month}' rather than the total downloads.
The ways of counting downloads can vary for different models.
Besides, according to the explanation of how Hugging Face collects model download statistics,\footnote{\url{https://huggingface.co/docs/hub/en/models-download-stats}} by default every HTTP request the \texttt{config.json} file in a repository is counted as a download.
However, this does not necessarily mean that the model is fully downloaded and used by the user.
Thus, we opt to only use the number of likes as the popularity measurement. 
Hugging Face users can contribute to the ecosystem by uploading their models and datasets. 
We calculate the total number of likes of all models owned by a user.
A higher number of likes indicates that the models contributed by the user are well-received and widely appreciated by the community, reflecting the user's influence and popularity in the ecosystem.

\vspace*{0.2cm}
\noindent \textbf{Dependency Network Analysis.}
In the LLM4Code ecosystem, models and datasets are interconnected through dependencies, forming a complex network.
Take an example from Figure~\ref{fig:example}, the model \path{codeparrot/codeparrot-small-code-to-text} is fine-tuned from \path{codeparrot/codeparrot-small} on the dataset \path{codeparrot/github-jupyter-code-to-text}. It indicates one model-to-model dependency and one model-to-data dependency.
By manually labeling the dependencies in the ecosystem (as described in Section~\ref{subsec:step2}), we construct three networks that capture the relationships between models and datasets.
The networks are: owners (O)-network, models (M)-network, and datasets (D)-network.
O-network only considers the dependencies between owners and datasets/models. 
M-network and D-network only consider the dependencies between models and dependencies between datasets, respectively. 
Then, we analyze network properties to understand the structure and evolution of LLM4Code.
Specifically, we compute the degree of each node (i.e., the number of edges connected to it) and then the degree distribution~\cite{wasserman_faust_1994}.
Degree distribution can help us understand the topology of the network, e.g., whether there exist a few highly connected nodes that demonstrate strong influence over the network. 
It also helps estimate the growth mechanisms of the ecosystem~\cite{barabasi1999emergence}.

\subsubsection{Results}

We rank models by their number of likes.
The full rankings are available in our online replication package, and Table~\ref{tab:degree} presents the top five.
The top-5 most liked LLM4Code are \path{bigcode/starcoder}, \path{replit/replit-code-v1-3b}, \path{bigcode/santacoder}, \path{microsoft/codebert-base}, and \path{Salesforce/codegen-16B-mono}.
We notice that these models are all owned by companies, which motivates us to understand the difference between contribution from company accounts and non-company accounts. 
We formulate three hypotheses:

\vspace*{0.1cm}
\begin{quote}
\hangindent=0.5cm
\hangafter=1
\textbf{Hypothesis 1: }\textit{Models owned by companies are more popular than that owned by non-company accounts.}
\end{quote}

\begin{quote}
  \hangindent=0.5cm
  \hangafter=1
  \textbf{Hypothesis 2: }\textit{Company accounts contribute more models than non-company accounts.}
  \end{quote}

\begin{quote}
\hangindent=0.5cm
\hangafter=1
\textbf{Hypothesis 3: }\textit{Company accounts receive more likes than non-company accounts.}
\end{quote}
\vspace*{0.1cm}

To test these hypotheses, we divide the models into two groups: company-owned models and non-company-owned models, based on the metadata of the model owners. 
Hugging Face uses a `\texttt{Company}' or `\texttt{Enterprise}' tag to indicate that the owner is a company.
We calculate the average likes for each group, finding that company-owned models receive 97.42 likes — 46.5 times higher than non-company-owned models, which has 2.05 likes on average.
A Mann-Whitney U Test is conducted and shows that the difference is statistically significant ($p<0.01$), with a large Cliff's Delta effect size ($\delta=0.63$), supporting Hypothesis 1.
However, the number of models contributed by the two groups does not differ significantly ($p=0.51, \delta=0.12$), which leads us to reject Hypothesis 2
We further calculate the total number of likes for each owner, with the top-5: \texttt{bigcode}, \texttt{Salesforce}, \texttt{replit}, \texttt{microsoft}, and \texttt{facebook}.
Owners are categorized into company accounts and non-company accounts.
We conduct a Mann-Whitney U Test to show company accounts receives significantly more likes than non-company accounts ($p<0.01, \delta=0.72$).

\begin{boxA}
  \textbf{Finding 1.}
  Company and non-company users contribute a similar average number of models, but models from company users receive significantly more likes than those from non-company users, showing their higher impact in the ecosystem.
\end{boxA}

We then analyze importance of models and datasets in the ecosystem from a network perspective.
We analyze the degree distribution of O-network, M-network, and D-network. 
We use \texttt{powerlaw} package and show that degree distributions follow a \textit{power-law distribution} ($p<0.01$).
This suggests that a few nodes, often called “hubs,” have a much higher degree than average.
The power-law pattern typically reflects the \textit{preferential attachment mechanism}~\cite{barabasi1999emergence}, where new nodes are more likely to connect to those with many existing links. 
This may be because users tend to reuse LLM4Code models and datasets that are already popular. 
High-degree nodes can serve as valuable entry points for new users to contribute to the ecosystem.

The out-degree of a node is the number of edges starting from this node, indicating the number of models or datasets derived from this node.
In M-network, the top-5 mostly reused models to build new ones are: \path{t5-small}, \path{bigcode/santacoder}, \path{t5-base}, \path{Salesforce/codet5-base}, and \path{bigcode/starcoder}.
We are curious the relationship between the out-degree (i.e., a model's importance to the ecosystem expansion) and the number of likes (i.e., users' recognition of a model). 
A Spearman correlation test shows a weak positive correlation (coefficient $0.22$, $p<0.01$) between out-degree and the number of likes. 
This suggests that the community's awareness may not align with the practical significance of LLM4Code. 
For instance, when developing a new LLM4Code, users may benefit from exploring models with more dependents, rather than simply starting with the most liked models. 
This highlights the importance of analyzing LLM4Code with a focus on dependencies.

\begin{boxA}
  \textbf{Finding 2.}
  The popularity of a model or dataset does not always align with its importance in the ecosystem. Simply starting with the most liked models may not be the best strategy for developing new LLM4Code.
\end{boxA}

\begin{table*}[!t]
  \centering
  \caption{The categorization of reuse types in the code model ecosystem. Each model is labelled with the \textit{closest type} of reuse. For example, a model that is fine-tuned from another model is labelled as \textit{Fine-tune} rather than \textit{Architecture}.}
  \label{tab:reuse-types}
  \small
  \begin{tabular}{lp{9.5cm}c}
    \toprule
    \textbf{Type} & \textbf{Explanation} & \textbf{Count} \\
    \midrule
    \rowcolor{gray!15} \textbf{Fine-tune} & Model A fine-tunes Model B on other datasets. For example, \texttt{\path{usvsnsp/code-vs-nl}} fine-tunes \texttt{\path{distilbert-base-uncased}} on \texttt{\path{codeparrot/github-code}}. &  205 \\
    \textbf{Architecture} & Model A shares the same architecture as Model B. For example, \texttt{Salesforce/codet5-base} shares the architecture of \texttt{t5-base}. & 84 \\
    \rowcolor{gray!15} \textbf{Quantize} & Model A is a quantized version of Model B. For example, \texttt{\path{mayank31398/starcoderbase-GPTQ-4bit-128g}} is quantized using 4-bit from \texttt{\path{bigcode/starcoder}}. & 32 \\
    \textbf{Continue} & Model A uses the parameters of Model B and continue to pre-train on other datasets. For example, \texttt{\path{neulab/codebert-c}} uses \texttt{\path{microsoft/codebert-base-mlm}} and continues to pre-train on the \texttt{\path{codeparrot/github-code-clean}} dataset using the masked-language-modeling task. & 15 \\
    \rowcolor{gray!15} \textbf{Model Conversion} & Model A is a converted version from Model B using a converter. For example, \texttt{\path{SirWaffle/codegen-350M-multi-onnx}} is obtained by converting \texttt{\path{Salesforce/codegen-350M-multi}} into the ONNX format. & 14 \\
    \textbf{Distillation} & Model A is obtained by knowledge distillation from Model B. For example, \texttt{\path{jiekeshi/CodeBERT-3MB-Clone-Detection}} is fine-tuned using the labels generated by \texttt{\path{microsoft/codebert-base}}.  & 12 \\
    \rowcolor{gray!15}  \textbf{Adapter} & Model A is obtained by adding pairs of rank-decomposition weight matrices to existing weights of Model B. For example, \texttt{\path{kaiokendev/SuperCOT-LoRA}} contains the Lora weight trained on the \texttt{\path{neulab/conala}} dataset, which is compatible with \texttt{\path{LLama}} models. & 9 \\
     \textbf{Instruction-tune} & Model A fine-tunes Model B on a collection of tasks described using instructions. For example, \path{sambanovasystems/starcoder-toolbench} is instruction-tuned & 9 \\
     \rowcolor{gray!15} \textbf{Adversarial} & Model A is obtained by adversarially training Model B on a dataset. For example, \texttt{\path{jiekeshi/CodeBERT-Adversarial-Finetuned-Clone-Detection}} is an adversarially trained from \texttt{\path{microsoft/codebert-base}}. & 6  \\
    \textbf{Unclear} & The documentation of a model provides insufficient information to decide the nature of its relationship with other models. & 49 \\
    \bottomrule
  \end{tabular}
\end{table*}

\subsection{RQ2. What are the practices of reusing LLM4Code?}
\label{subsec:rq2}
\subsubsection{Motivation}

A powerful language model is usually pre-trained on large datasets using significant computational resources to acquire general knowledge. 
Subsequently, users can reuse this model to various downstream tasks with much less computing power.
This is similar to software reuse in the open-source community, where developers build on existing frameworks and packages instead of starting from scratch. However, model reuse practices in LLM4Code are not well-explored in current literature. 
Questions like the types of reuse and their distribution remain unanswered. Understanding these practices can illuminate user needs and preferences, guiding new model development. 
For example, if users aim to develop a model running on resource-constrained devices, they can refer to LLM4Code reused by \textit{quantization} in the ecosystem and learn from their practices.

\subsubsection{Methodology}
Building on previous research in software content categorization~\cite{10.1145/2568225.2568233,10.1145/1718918.1718973,yang2023users}, we employ open card sorting to classify the dependencies between LLM4Code  (i.e., how a model is reused) into distinct categories.
Our objective is to identify the \textit{closest dependency}, which refers to the most immediate and significant relationship between a reused model and its base model.
For example, let us consider a Model A \path{stmnk/codet5-small-code-summarization-python} that is obtained by finetuning Model B \path{Salesforce/codet5-small}, and the latter shares the same architecture as Model C \path{t5-small}.
We call finetuning Model B the closest dependency of Model A.
As we also know the dependency between Model B and Model C, we can infer that Model A has an architecture sharing dependency on Model C.
The card sorting process to build the model reuse taxonomy is as follows:

\begin{enumerate}[leftmargin=*]
	\item \textit{Preparation}: Following the approach described in Section~\ref{subsec:step2}, we read the model cards of all LLM4Code to understand whether a dependency relationship exists between two models. 
  We collect $435$ pairs of models that have dependencies. 
	\item \textit{Execution}: Two authors of this paper discuss with each other to categorize the dependencies into meaningful groups. 
  \textit{Open card} sorting is adopted, which assumes no pre-defined groups and lets the groups emerge and evolve during the sorting process. 
  In this step, the two authors are encouraged to create groups. 
  This step takes a total of around 10 hours.
	\item \textit{Analysis}: After identifying groups in the previous step, another expert with LLM4Code background joins the discussion to merge groups of relevant topics into more general categories. 
  This step takes roughly 2 hours. 
\end{enumerate}
After constructing the taxonomy, we explain each model reuse type and report the distribution of different types to understand the practices in reusing LLM4Code.

\subsubsection{Results}
The open card sorting process identifies 9 types of model reuse. 
Dependencies that cannot be categorized (e.g., when the model card lacks relevant information) are labeled as `\textit{unclear}.'
Table~\ref{tab:reuse-types} shows the distribution of the dependency types, sorted by the number of occurrences.
Below, we explain each type of model reuse and their examples in the same order as Table~\ref{tab:reuse-types}.

\vspace*{0.2cm}
\noindent \textbf{Fine-tune.} Fine-tuning involves taking a pre-trained model (usually trained on a large dataset) and adapting it to a new, often related task by further training on a smaller, specific dataset. 
This reuse offers faster training times and improved performance, especially when the new dataset is small and insufficient for training a large model from scratch.
\path{usvsnsp/code-vs-nl} fine-tunes \path{distilbert-base-uncased} on \path{codeparrot/github-code}. 
Fine-tuning is the most common method for reusing LLM4Code, accounting for 205 out of 435 dependencies (47.13\%).

\vspace*{0.2cm}
\noindent \textbf{Architecture.}
Architecture reuse involves using the same architecture as another model but training from scratch rather than fine-tuning. For example, \path{Salesforce/codet5-base} shares the architecture of \path{t5-base} but is trained from scratch without using the latter's pre-trained weights. This form of reuse accounts for 84 out of 435 dependencies (19.31\%).

\vspace*{0.2cm}
\noindent \textbf{Quantize.}
Quantization reduces the precision of a model's weights and biases, typically converting floating-point numbers to lower-bit-width integers. 
Its main goal is to lower memory usage and computational demands, enabling deployment in resource-constrained environments like mobile or edge devices. 
For example, \path{mayank31398/starcoderbase-GPTQ-4bit-128g} is quantized to 4-bit from \path{bigcode/starcoder}, reducing the model size from 63 GB to 9 GB. 
This quantization accounts for 32 out of 435 dependencies (7.36\%), ranking as the third most common reuse type.
It aligns with the recent research trend of producing more efficient LLM4Code~\cite{wei2023greener,shi2024efficient,avatar}.

\vspace*{0.2cm}
\noindent \textbf{Continue Training.} 
Continue training falls between fine-tuning and architecture sharing. 
It involves using the parameters of an existing model and continuing to pre-train on other datasets. 
For example, \path{neulab/codebert-c} builds on \path{microsoft/codebert-base-mlm} by continuing pre-training on the \path{codeparrot/github-code-clean} dataset using the masked-language-modeling task. 
This reuse accounts for 15 out of 435 dependencies (3.45\%).

\vspace*{0.2cm}
\noindent \textbf{Model Conversion.} 
Model conversion involves transforming LLM4Code from one framework or format to another. 
This is often necessary due to differences between the training and deployment environments. 
For instance, a model trained in TensorFlow may need to be deployed using ONNX Runtime for optimized inference. 
Conversion also enables hardware acceleration; to leverage Apple's neural engine, models must be converted to Core ML. 
Examples of such conversions include \path{SirWaffle/codegen-350M-multi-onnx}, converted from \path{Salesforce/codegen-350M-multi} to ONNX, and \path{rustformers/mpt-7b-ggml}, a GGML version of \path{mosaicml/mpt-7b}. These conversions account for 14 out of 435 dependencies (3.22\%).

\vspace*{0.2cm}
\noindent \textbf{Distillation.}
Distillation, or \textit{knowledge distillation}, transfers knowledge from a larger, complex 'teacher' model to a smaller 'student' model. 
This technique enables deploying smaller, faster models with lower memory requirements, ideal for resource-constrained environments like mobile devices~\cite{compressor}. For example, \texttt{\path{jiekeshi/CodeBERT-3MB-Clone-Detection}} is distilled from \texttt{\path{microsoft/codebert-base}}. 
Such distillation accounts for 12 out of 435 dependencies (2.76\%).

\vspace*{0.2cm}
\noindent \textbf{Adapter.} 
An adapter is a small set of additional layers or modules inserted into LLM4Code to adapt it for a new task or domain. 
One example is Low-Rank Adaptation (LoRA)~\cite{hu2021lora}, which accelerates LLM4Code training by updating only a subset of parameters. 
It achieves this by adding rank-decomposition weight matrices (update matrices) to existing weights and training only the new matrices. 
For instance, \texttt{\path{kaiokendev/SuperCOT-LoRA}} contains LoRA weights trained on the \texttt{\path{neulab/conala}} dataset and is compatible with \texttt{\path{LLama}} models. 
This type of reuse accounts for 9 of 435 dependencies (2.07\%).

\vspace*{0.2cm}
\noindent \textbf{Instruction-tuning.}
Instruction-tuning refers to the practice of fine-tuning a model on a collection of tasks described using instructions.
As a new parameter-efficient training approach, we separate it from the fine-tuning category.
An instance is that \texttt{\path{teknium/Replit-v1-CodeInstruct-3B}} is instruction-tuned on the \texttt{\path{replit/replit-code-v1-3b}} model.
This type of reuse accounts for 9 out of 435 dependencies (2.07\%).

\vspace*{0.2cm}
\noindent \textbf{Adversarial training.} 
It is used to enhance model robustness by generating adversarial examples and augmenting the training data with them. 
For example, \texttt{\path{jiekeshi/CodeBERT-Adversarial-Finetuned-Clone}} is adversarially trained from \texttt{\path{microsoft/codebert-base}}. 
Despite growing interest in the software engineering community~\cite{alert,coda,Yefet2020}, adversarial training is rare in the code model ecosystem, with only 6 instances out of 435 dependencies (1.38\%).

\begin{boxA}
  \textbf{Finding 3.}
  We categorize model reuse into 9 types, with fine-tuning being the most common, followed by architecture sharing. We also observe several types of reuse aimed at improving the \textbf{efficiency in training and deployment of LLM4Code}. Additionally, there is some interest in supporting \textbf{cross-platform and cross-framework compatibility} and enhancing \textbf{LLM4Code robustness}, but these efforts are relatively scattered.
\end{boxA}

\begin{figure*}[!t]
  \centering
  \includegraphics[width=1\textwidth]{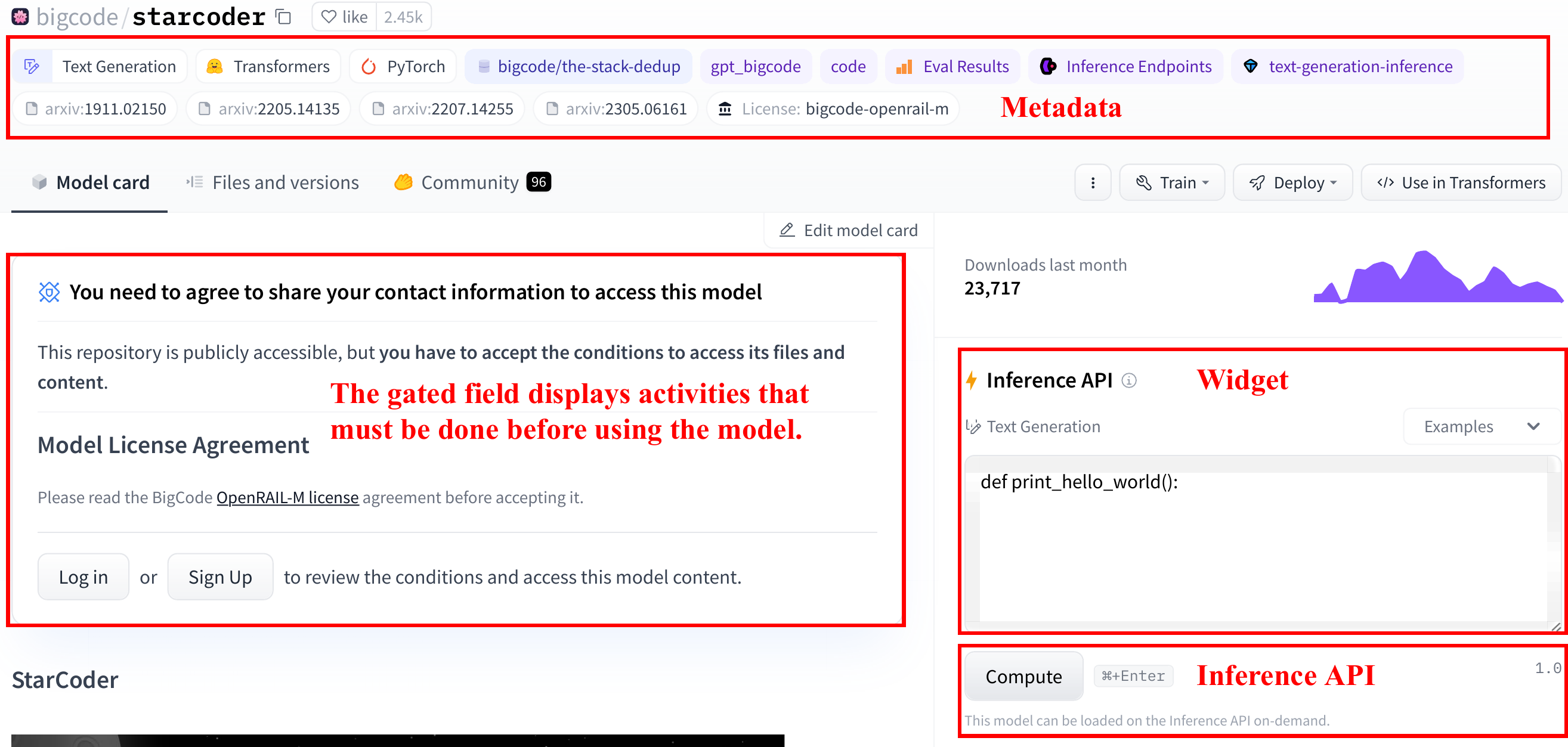}
  \caption{The rendered model card of \texttt{bigcode/starcoder} and features provided by Hugging Face, e.g., gated field, widget, and inference API.}
  \label{fig:demo}
\end{figure*}

\subsection{RQ3. What are the practices of publishing LLM4Code?}

\subsubsection{Motivation.}
After developing LLM4Code, users can publish models to Hugging Face, contributing to the ecosystem. 
This allows others to access and build upon these models, fostering prosperity within the ecosystem. 
Thus, model publishing practices are key to the evolution and success of LLM4Code. 
Taking the documentation practice as an example. 
Model documentation confirms whether a model belongs to the ecosystem and clarifies its reuse potential with other models. 
Without proper documentation, even high-quality models may be underutilized, limiting their contribution. 
This highlights the need to examine current publishing practices and identify pain points in the LLM4Code ecosystem.

\subsubsection{Methodology.}
In this research question, we focus on two important aspects of practices when publishing LLM4Code: (1) documentation practice and (2) license practice.

\vspace*{0.2cm}
\noindent \textbf{Documentation Practice.}
Good documentation in open-source software and models is crucial~\cite{aggarwal2014co} as it ensures accessibility and usability, allowing developers of different skill levels to understand, use, and contribute to the project.
Fan et al.~\cite{fan2021makes} find that some documentation practices (e.g., the use of lists, code blocks, and images, etc) affect popularity of AI repositories on GitHub.
Thus, we consider LLM4Code documentation to understand the usage of lists, code blocks, inline code, images, tables, and external links. 
We download the README.md files of LLM4Code and use regular expressions to extract the information.
For example, Markdown supports both ordered and unordered lists.
Unordered lists usually start with \texttt{*}, \texttt{-}, or \texttt{+}, and ordered lists usually start with a number followed by a period (e.g., \texttt{1.}).
We use the regular expression `\texttt{(?:\^{}|$\backslash$n)(?!$\backslash$s*[-*]$\backslash$s*$\backslash$n)$\backslash$s*[-*]$\backslash$s}' to detect unordered lists in the documentation.
Our replication package includes the scripts to extract the relevant information from the documentation.
We also compare the documentation practice of LLM4Code with that of AI repositories hosted on GitHub~\cite{fan2021makes}.

\vspace*{0.2cm}
\noindent \textbf{License Practice.}
A software license is a legal agreement defining the terms for using, modifying, or distributing software. 
It provides a framework for users to understand their rights and limitations. 
As LLM4Code models gain popularity and are deployed in real-world applications, understanding licensing practices within code model ecosystems is crucial.
We extract what licenses are used in LLM4Code by reading their documentation.
We compare the license usage (including both types of license and distribution of license) in LLM4Code ecosystems with that in public software
source code (using the dataset of Software Heritage~\cite{SoftwareHeritage}) to understand the difference.
Additionally, we use the Linux Foundation's OSS license compatibility table~\cite{licensetable}, which defines a list of incompatible license usages given a license, to analyze the potential license compatibility issues in each pair of LLM4Code with dependencies.

\subsubsection{Results}

\vspace*{0.2cm}
\noindent \textbf{Documentation Practice}
The documentation of a Hugging Face model has two parts: the \textit{metadata} written in YAML and the \textit{description} written in Markdown.
The metadata is used to render the model card on the Hugging Face website, as shown in Figure~\ref{fig:demo}.
We find that 40 out of 366 (10.93\%) LLM4Code do not provide any metadata.
The metadata contains different information, and we find 17 unique fields in the LLM4Code's metadata.
For example, 153 models specify the \textit{license} field, 68 models provide the \textit{datasets} field, and 2 models provide the \textit{base\_model}.
Although the metadata does provide some information of the model's dependencies, it is hard to automatically build the ecosystem purely relying on the metadata.
For example, only 2 models provide the base model information, while we manually identify 435 model dependencies, demonstrating the value and necessity of manual effort to understand the ecosystem.

\begin{table}[!t]
  \centering
  \caption{The average frequency of different types of information in LLM4Code documentation within our ecosystem is compared to that of general AI repositories on GitHub. We compute this by dividing the total occurrences by the number of documentation files. Cliff's $\delta$ and $p$-values are reported to assess effect size and statistical significance.}
  \label{tab:feature_metrics}
  \begin{tabular}{lrrrr} %
    \toprule
    \textbf{Information} & \textbf{LLM4Code} & \textbf{AI Repos} & \textbf{Cliff's $\delta$} & \textbf{$p$-values} \\ %
    \midrule
    \rowcolor{gray!15} \textbf{Lists}        & 3.77  & 6.02  &  -0.27  & $<0.01$ \\ %
    \textbf{Code Blocks}  & 1.25  & 3.92  &  -0.22  & $<0.01$ \\
    \rowcolor{gray!15} \textbf{Inline Code}   & 11.03 & 38.63 &  -0.33  & $<0.01$ \\
    \textbf{Tables}       & 0.55  & 0.13  &  +0.02  & $>0.05$ \\
    \rowcolor{gray!15} \textbf{Images}        & 0.07  & 0.53  &  -0.19  & $<0.01$ \\
    \textbf{URLs}         & 5.90  & 8.46  &  -0.13  & $<0.01$ \\
    \bottomrule
  \end{tabular}
\end{table}

Some metadata is relevant to advanced Hugging Face features.
For example, \textit{widget} is used to create a widget\footnote{\url{https://huggingface.co/docs/hub/en/models-widgets}} that allows users to run inferences directly in model webpage, as shown in the bottom right part of Figure~\ref{fig:demo}.
There are 169 models use the widget function while only 17 models use the inference API.
Model owners can specify \textit{extra\_gated\_prompt} and \textit{extra\_gated\_fields}, which will render a \textit{gated field} (as shown in Figure~\ref{fig:demo}, bottom-left), asking users to accept the terms and conditions in order to access the model. 
However, this feature is rarely used, with only 7 models (2\%) adopting it.

We then analyze the \textit{description} part in the documentation.
Fan et al.~\cite{fan2021makes} defines a set of features in the README.md file that shows the documentation quality: the average numbers of lists, code blocks, inline code, tables, images, and URLs in each description.
We compare the feature differences between LLM4Code and a collection of 1,149 AI repositories hosted on GitHub, which is collected by Fan et al.~\cite{fan2021makes}.
The comparison is shown in Table~\ref{tab:feature_metrics}, accompanied by the effect size and statistical significance.
We can observe that AI repositories on GitHub tend to include more information (including lists, code blocks, inline code, images, and URLs) in the README.md file than LLM4Code.
However, LLM4Code tend to include more tables (0.55 tables per model) than the GitHub AI repositories do (0.13 tables per repository), but the difference is not statistically significant ($p$-value $>0.05$).

\begin{boxA}
  \textbf{Finding~4.}
  The metadata of LLM4Code is not enough to automatically build the ecosystem, and the documentation of LLM4Code is generally less informative than documentation of AI repositories on GitHub, calling for more attention on improving the documentation quality.
\end{boxA}

\noindent \textbf{License Practice}
Out of 366 models, 211 LLM4Code lack any license information in their metadata or description.
For those with license, the most commonly used licenses are Apache-2.0 (33.33\%),  BSD-3-Clause (30.0\%), bigcode-openrail-m (16.0\%), MIT (10.6\%), and openrail (4.0\%).
This distribution shows a different pattern compared to traditional software repositories.
As shown by Gonzalez-Barahona et al.~\cite{Jesus2023}, in Software Heritage, which contains more than 11 billion public software artifacts, the most commonly used licenses are MIT (39.3\%), GPL (13.4\%), BSD-3-Clause (7.3\%) and Apache-2.0 (5.5\%).
LLM4Code tend to take less restrictive licenses than traditional software repositories, as evident by the low usage of GPL licenses.
Some LLM4Code adopt AI-specific licenses.
For example, 21.10\% of models use RAIL (Responsible AI Licenses) or its variants.
RAIL stands for Responsible AI Licenses, a set of licenses designed to promote responsible AI research and development.
To be more specific, 32, 13 and 1 models use \path{bigcode-openrail-m}, \path{openrail}, and \path{bigscience-bloom-rail-1.0}. 
Another AI-specific license is the AI model license agreement by AI2,\footnote{\url{https://allenai.org/impact-license}} used by 8 models in the ecosystem.

We further investigate the potential license incompatibility issues given a model reuse.
We use the Linux Foundation's OSS license compatibility table~\cite{licensetable}, which defines a list of incompatible license usages for a license.
Note that we find 7 models that use more than one license.
For example, \path{allenai/open-instruct-code-alpaca-7b} mentions that it is licensed under both AI model license agreement and the original Llama license.\footnote{\url{https://github.com/facebookresearch/llama/blob/main/LICENSE}}
We exclude them from the analysis.
Our analysis reveals no instance of license incompatibility in the current ecosystem.
The results are as expected because the specified licenses are generally less restrictive (e.g., Apache-2.0, MIT, etc) compared to strict licenses like GPL.
Besides, as we mentioned earlier, over 60\% of models do not provide any license information, even for some popular ones like \path{Salesforce/codet5-base}. 
This may lead to risks that when the upstream models update their license information, many models in the ecosystem will be affected.
So we still need to be cautious about the potential license issues in the ecosystem.

\begin{boxA}
  \textbf{Finding~5.}
  LLM4Code typically uses less restrictive licenses than traditional software repositories, with Apache-2.0, RAIL, BSD-3-Clause, and MIT being the most common. While no license incompatibility issues have been identified in the current ecosystem, potential risks remain and warrant attention.
\end{boxA}

\section{Automated Data Collection} \label{sec:automation}

The LLM4Code ecosystem is growing rapidly. 
On August 13, 2023, searching using the keyword `code' on Hugging Face returns a 1,566 models; by September 6, 2024, this number had grown to 10,324. 
Manual labeling to build and scale this ecosystem is time-consuming and costly. 
Recent research shows that LLMs can achieve inter-rater agreements comparable to human raters, reducing the need for human annotation~\cite{ahmed2024llmsreplacemanualannotation}.
Therefore, we explore the potential of LLMs to automate the LLM4Code ecosystem construction at scale.

\subsection{Model Selection}

Different LLMs vary in their ability to accurately annotate data, as well as in cost.
For example, GPT-4o is generally considered providing more accurate answers than its smaller version GPT-4o mini, but the former is more expensive to use.\footnote{At the time of writing the paper, GPT-4o costs \$5 per million input tokens and GPT-4o mini costs \$0.15 per million input tokens. Details can be found at \url{https://openai.com/pricing}}
Given the strong performance demonstrated by OpenAI models in various studies, we utilize their APIs, including GPT-4o, GPT-4o mini, and GPT-3.5 Turbo, for data annotation.

\begin{figure*}[!t]
  \centering
  \includegraphics[width=0.9\textwidth]{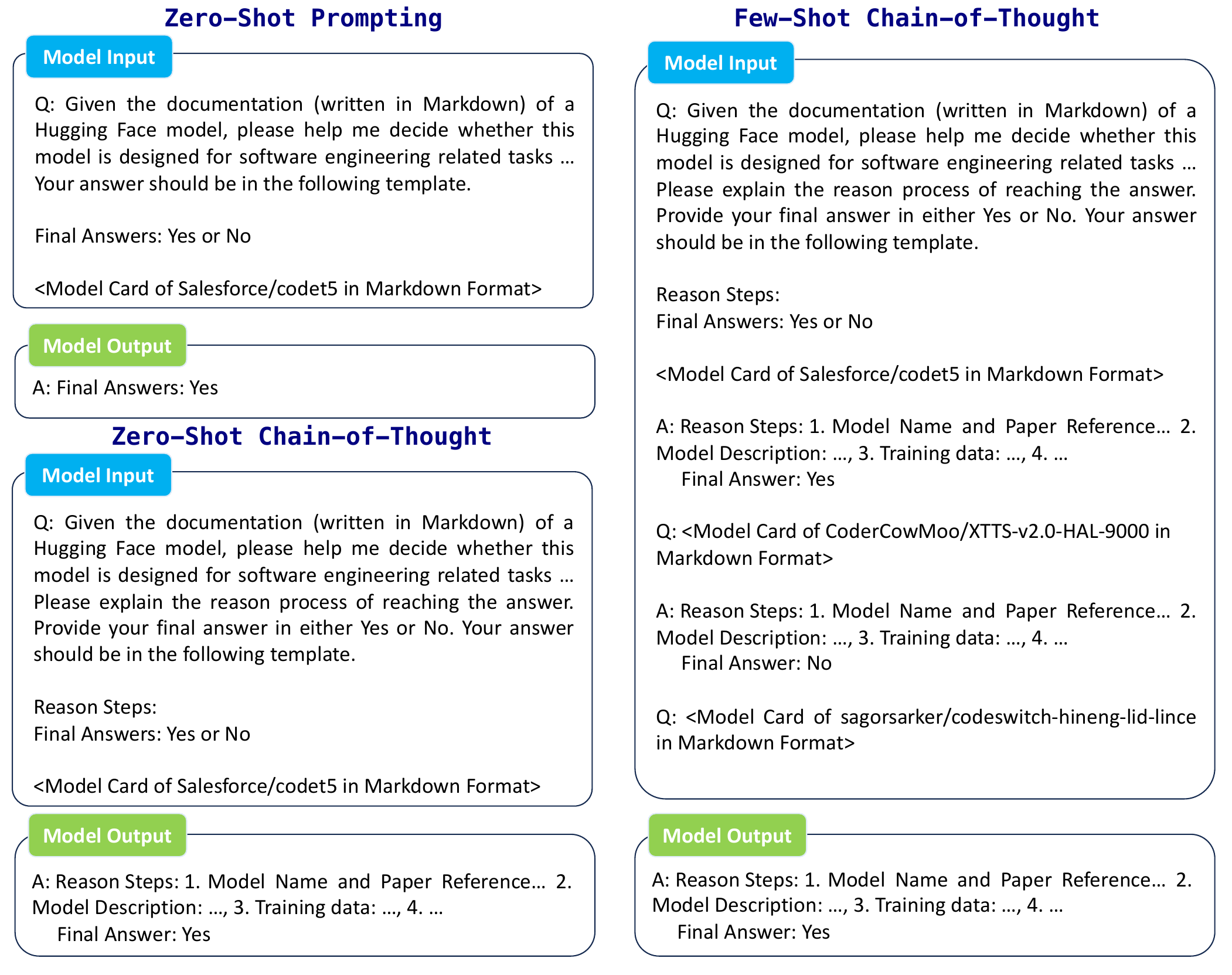}
  \caption{Three prompting strategies to let LLMs infer whether a model is LLM4Code given its documentation.}
  \label{fig:prompt}
\end{figure*}

\subsection{Tasks and Prompt Engineering}
We evaluate the ability of LLMs to automate three tasks in ecosystem construction: (1) \textit{infer model type}—determine if a model is LLM4Code based on its model card, (2) \textit{infer base models}—identify the models used to derive a given model, and (3) \textit{infer dependency types}—determine how base models are reused in model derivation.

Prompts are crucial to guide LLMs to produce high-quality responses.
We experiment with three prompting strategies: (1) \textit{zero-shot prompting} (ZS-P), (2) \textit{zero-shot chain-of-thought} (ZS-CoT), and (3) \textit{few-shot chain-of-thought} (FS-CoT). In zero-shot prompting, we provide the model's documentation and directly ask it to infer information. The LLM is instructed to read the documentation of a Hugging Face model and respond with a simple 'Yes' or 'No,' ensuring clear, easily evaluated answers.

Researchers have found that asking models to provide intermediate reasoning steps, known as chain-of-thought (CoT) prompting~\cite{wei2022chain}, improves complex reasoning performance.
We explore two CoT strategies: zero-shot and few-shot CoT. 
In zero-shot CoT, the model is explicitly asked to provide reasoning steps for its answers. 
In few-shot CoT, we additionally provide two examples of the reasoning steps. 
Specifically, we manually construct reasoning steps to explain why `\path{Salesforce/codet5p-220m}' is an LLM4Code and why `\path{CoderCowMoo/XTTS-v2.0-HAL-9000}' is not.\footnote{These models are randomly selected from our dataset, but others could be used as well.}
Simplified prompts and the expected LLM outputs, are shown in Figure~\ref{fig:prompt}.
Due to the space limits, we put the prompts for other tasks in our replication package.

\subsection{Primary Results Analysis}

We evaluate three LLMs on the manually labeled models with available documentation (as described in Section~\ref{subsec:step1}). 
Since \path{Salesforce/codet5p-220m} and \path{CoderCowMoo/XTTS-v2.0-HAL-9000} are used as examples in FS-CoT, they are excluded from the evaluation. 
We report each LLM's accuracy on the three tasks, calculated as the percentage of answers matching the manual labels.

\subsubsection{Model Type Inference}

\begin{table}[!t]
  \caption{Three LLMs' performance in inferring model types using different prompting. `Acc' refers to the accuracy and `F1' refers to the F1-score.}
  \centering
  \label{tab:model_performance}
  \begin{tabular}{lcccccc}
    \toprule
    & \multicolumn{2}{c}{ZS-P} & \multicolumn{2}{c}{ZS-CoT} & \multicolumn{2}{c}{FS-CoT} \\
    \cmidrule(lr){2-3} \cmidrule(lr){4-5} \cmidrule(lr){6-7}
    & Acc & F1 & Acc & F1 & Acc & F1 \\
    \midrule
    \rowcolor{gray!15} GPT-4o & 0.931 & 0.955 & 0.882 & 0.922 &  0.980 & 0.987 \\
    GPT-4o mini  & 0.980  & 0.988 & 0.867 & 0.911 & 0.963 & 0.981 \\
    \rowcolor{gray!15} GPT-3.5-turbo & 0.899 & 0.932 & 0.867 & 0.910 & 0.815 & 0.872 \\
    \bottomrule
  \end{tabular}
\end{table}

All three LLMs can follow the prompt well to provide their answers in the required format, i.e., putting the answer `\texttt{yes}' or '\texttt{no}' at the end of the response after the keyword `\texttt{answers: }'.
Thus, we only extract the content after the keyword `\texttt{answers: }' to determine the model type.
If only `\texttt{yes}' appears, the model is considered as an LLM4Code, otherwise it is not.
We compare the labels generated by the LLMs with the manually annotated labels. 
The results are summarized in Table~\ref{tab:model_performance}.
We notice that GPT-4o-mini achieves desirable performance (accuracy of 0.980 and F1 of 0.988) when using the zero-shot prompt.
CoT does not lead to significant improvement, likely because the task is simple and does not require complex reasoning.
Considering the performance and cost, GPT-4o mini using the zero-shot prompt is an appropriate choice in this task.

\subsubsection{Base Model Inference}

Unlike model type inference, this task requires the LLM to output a model name rather than a binary answer.
We find it difficult to automatically analyze whether the base model is correctly inferred.
By reviewing the LLM outputs, we identify two main reasons. 
One reason is that LLMs may use different expressions for the same model.
A Hugging Face model's name follows the format of \texttt{owner/model-name}, e.g., \texttt{Salesforce/codet5p-220m}. 
However, LLMs may express the model name as `\texttt{codet5p-220m by Salesforce}' or `\texttt{Salesforce codet5p-220m}', which both refer to \texttt{Salesforce/codet5p-220m}.
Such variety in expressions makes it challenging to automatically decide whether the model name is correctly inferred by string match.
Another reason is that the inferred base model may not share the closest relationship (defined in Section~\ref{subsec:rq2}).
For example, \path{SEBIS/code_trans_t5_base_api_generation_multitask_finetune} is derived by fine-tuning \path{SEBIS/code_trans_t5_base_api_generation_multitask}.
However, LLMs identify \path{t5-base} as the base model, which is also correct as the model shares the same architecture with the \path{t5-base}.
However, in our manually labelled data, we record the closest relationship and finetuning is considered as a closer relationship than sharing the same architecture.

To evaluate LLMs' performance in this task, we address the first issue by manually converting the LLM-inferred model names (of different expressions) to the standard format.
The inference is considered correct if LLMs provide the correct but not closestly related base model, as it also reveals the model derivation information.
We present the results in Table~\ref{tab:base_model_inference}.
GPT-4o performs best in this task but its smaller version GPT-4o mini also achieves a good performance, only 0.012 lower than GPT-4o in terms of accuracy (with FS-CoT).
Although the three LLMs cannot fully automate this task, they can still provide valuable information to help us understand the base models of LLM4Code.

\begin{table}[!t]
  \caption{The accuracy of three LLMs in inferring the base model of a given LLM4Code using different prompts.}
  \centering
  \label{tab:base_model_inference}
  \begin{tabular}{lccc}
    \toprule
    & ZS-Prompt & ZS-CoT & FS-CoT \\
    \midrule
    \rowcolor{gray!15} GPT-4o & 0.844 & 0.866 & 0.878 \\
    GPT-4o-mini & 0.828 & 0.844 & 0.866 \\
    \rowcolor{gray!15} GPT-3.5-turbo & 0.838 & 0.838 & 0.844 \\
    \bottomrule
  \end{tabular}
\end{table}

\subsubsection{Dependency Type Inference}

Similar to the base model inference task, it is difficult to use string matching to automatically analyze whether the dependency type is correctly inferred, mainly due to that the LLMs do not follow the instructed template to organize the answer.
So to understand LLMs' performance, we manually compare the LLM outputs with the manually annotated dependencies.
The accuracy is shown in Table~\ref{tab:dependency_inference}.

We observe greater performance difference among LLMs.
GPT-4o outperforms GPT-4o-mini and GPT-3.5-turbo in by a large margin.
Specifically, when ZS-prompt is used, GPT-4o achieves an accuracy of 0.789, which is 0.287 higher than GPT-3.5-turbo and 0.230 higher than GPT-4o mini.
We see improvements brought by providing examples of steps to infer the dependency type.
The few-shot CoT strategy leads to 0.105, 0.124, and 0.150 higher accuracy than the ZS-prompt for GPT-4o, GPT-4o mini, and GPT-3.5-turbo, respectively.
It makes GPT-4o achieve a high accuracy of 0.894.
It still requires human intervention to decide the final results, but an accuracy of 0.894 demonstrates the promising potential to automate and assist the ecosystem construction.

\begin{boxA}
  \textbf{Finding~6.} LLMs can infer model types with high accuracy (0.98). For more complex tasks, such as base model and dependency type inference, they align well with human annotations, achieving accuracies of 0.878 and 0.894, respectively. Although human intervention is still needed, LLMs show potential for automating key tasks in constructing the LLM4Code ecosystem.
\end{boxA}

\begin{table}[!t]
  \caption{The accuracy of three LLMs in inferring the dependency type of a given LLM4Code using different prompts}
  \centering
  \label{tab:dependency_inference}
  \begin{tabular}{lccc}
    \toprule
    & ZS-Prompt & ZS-CoT & FS-CoT \\
    \midrule
    \rowcolor{gray!15} GPT-4o & 0.789 & 0.793 & 0.894 \\
    GPT-4o-mini & 0.559 & 0.595 & 0.683 \\
    \rowcolor{gray!15} GPT-3.5-turbo & 0.502 & 0.502 & 0.652 \\
    \bottomrule
  \end{tabular}
\end{table}

\subsection{Analysis of The Expanded LLM4Code Ecosystem}

The results show that LLMs can accurately determine whether a model is LLM4Code without human intervention. 
We use GPT-4o-mini to analyze all the 6,406 models with model cards (out of 10,324 models with `code' in their name collected in September 6, 2024).
GPT-4o-mini identifies 5,407 models as LLM4Code.
This process costs \$3.16 and 3 minutes in total.
In comparison, labelling 405 models manually costs two data annotators 11 and 13 hours.
It demonstrates the potential to scale up to the increasing ecosystem with a reasonable cost.

We fetch the LLM4Code metadata to understand the new trends and validate conclusions drawn from the manually labeled dataset.
We repeat the statistical tests to analyze Hypotheses 1, 2, and 3. 
We find that the conclusion drawn from the small manually labeled dataset still holds on the expanded dataset, confirming the importance of companies in the LLM4Code ecosystem.
Specifically, models owned by companies are statistically more popular (i.e., receiving more likes) than models owned by non-companies users ($p$-value $<0.01$).
Company accounts receive significantly more likes than non-company accounts ($p$-value $<0.01$) but they are similar in the number of uploaded models ($p$-value $>0.05$).
The community values the quality of the models and datasets more than the quantity.

\begin{figure}[!t]
  \centering
  \includegraphics[width=0.9\textwidth]{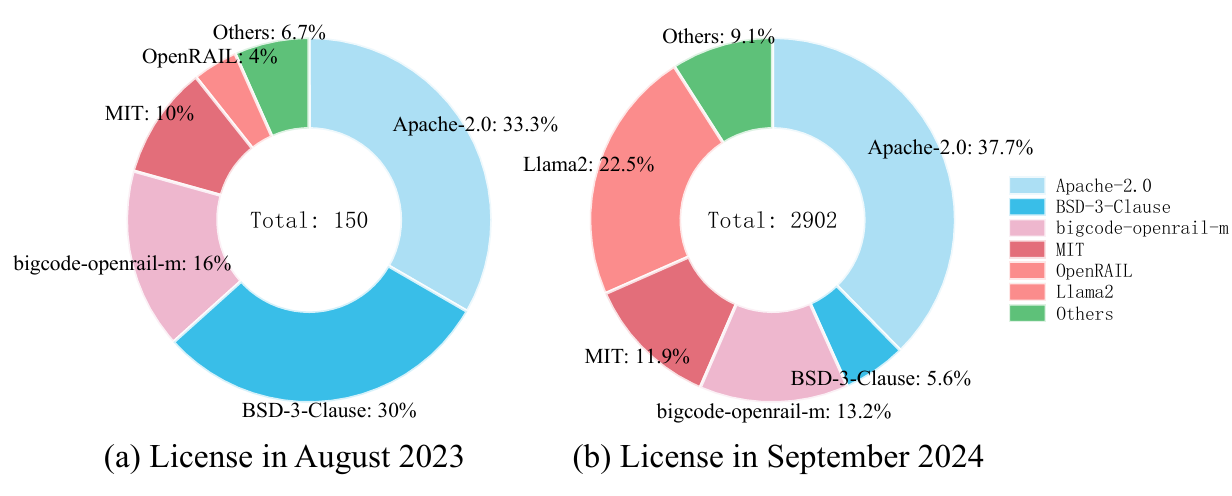}
  \caption{An illustrative example of the code model ecosystem, including code datasets, models and dependencies.}
  \label{fig:license}
\end{figure}

Out of 5,407 LLM4Code models, 3,143 lack license information in their metadata or are marked as `other', making up 58.12\%, close to the 57.66\% in the manually labeled dataset.
This indicates a persistent issue with missing license information during the development of LLM4Code ecosystem.
The top-5 most popular licenses are Apache-2.0 (37.66\%), Llama2 (22.54\%), bigcode-openrail-m (13.20\%), MIT (11.92\%), and BSD-3-Clause (5.62\%).
Figure~\ref{fig:license} shows the license distribution on the ecosystems conducted in August 2023 and September 2024, with popular licenses remaining less restrictive. 
The percentages of models Apache-2.0 and MIT licenses are overall stable, with slight increases of 4.4\% and 1.1\%, respectively.
But we notice the percentage of BSD-3-Clause license decreases by 16.8\%.
Regarding the AI-specific licenses, bigcode-openrail-m decreases by 2.8\%, while Llama2 surged by 22.5\%, indicating a trend towards more Llama-based models in the LLM4Code ecosystem.

\begin{boxA}
  \textbf{Finding~7.} 
  Some observations on the manually created ecosystem can generalize to the LLM-expanded one. We confirm the superiority of company-owned models in popularity and the persistent issue of missing license information. The ecosystem is dominated by less restrictive licenses. The increasing usage of Llama2 licenses suggests a trend towards more Llama-based LLM4Code.
\end{boxA}

\section{Discussion} \label{sec:discussion}

\subsection{Implications and Suggestions}
Our analysis of the LLM4Code ecosystem reveals important findings, summarized in Findings 1-7.
We derive implications from these findings and offer suggestions for stakeholders, such as the Hugging Face platform and LLM4Code developers, along with research opportunities.

\vspace*{0.2cm}
\noindent \textbf{Disparity between different contributors.}
Despite non-company users contributing a similar number of models as companies, company models receive significantly more likes, indicating higher impact (Finding 1). This disparity remains even with ecosystem expansion using LLMs (Finding 7), likely due to companies having more resources to train foundation models.
However, non-company users are crucial for diversifying models for various tasks, benefiting the community.
We suggest Hugging Face implement features to enhance visibility for high-quality models from non-company contributors.

\vspace*{0.2cm}
\noindent \textbf{Misalignment between popularity and importance.}
The misalignment between a model's popularity and its impact on deriving new models (Finding 2) can hinder users in selecting effective starting points for reuse. 
Hugging Face's `rank by likes' option might cause valuable, easily extendable models to be overlooked.
We suggest developers include base model information in model cards, and Hugging Face could add a feature to rank models by the number of derivatives. 
Researchers can also develop metrics and tools to better assess the contribution and usability of LLM4Code, aiding user decision-making.

\vspace*{0.2cm}
\noindent \textbf{Need for efficiency and compatibility.}
Users need efficient training and deployment when reusing models (Finding 3). The ecosystem also demands models that work across various platforms and frameworks, as well as robustness enhancement.
Researchers can explore innovative methods to improve training efficiency and robustness. Model developers should invest in efficient training techniques like quantization, distillation, and adapter models.

\vspace*{0.2cm}
\noindent \textbf{Insufficient Documentation and Metadata Quality.}
The metadata and documentation for LLM4Code models are often inadequate (Finding 4).
Compared to AI repositories on GitHub, LLM4Code documentation is less informative, potentially hindering usability and adoption~\cite{fan2021makes}. 
Hugging Face can promote better documentation by offering templates, checklists, and examples. Developers should prioritize clear, comprehensive documentation to ensure models are accessible to users of all expertise levels. Researchers can study how documentation quality affects model reuse and develop automated methods to improve it.

\vspace*{0.2cm}
\noindent \textbf{Licensing Practices and Potential Legal Risks.}
The prevalence of less restrictive licenses (Finding 5) suggests an ecosystem leaning toward openness.
However, missing license information poses legal risks and potential incompatibility issues. 
With the rise of AI-specific licenses (Finding 7), there is a need for better understanding and detection of legal risks of these licenses. 
We suggest Hugging Face to provide clear guidance on license selection.
LLM4Code developers and users should ensure models have appropriate, clear licensing to prevent legal issues and encourage responsible reuse.
Researchers can explore AI-specific licenses, develop tools to detect incompatibilities, and recommend suitable licenses for various scenarios.

\vspace*{0.2cm}
\noindent \textbf{LLMs for expanding the ecosystem.}
LLMs demonstrate high accuracy in inferring model types and show promise in automating key tasks within the ecosystem (Finding 6). However, for more complex tasks, human intervention remains necessary.
In a more general perspective, LLMs can infer metadata (e.g., base models) from the model documentation. 
Researchers could study how LLMs enhance model cards, and these tools could be integrated into the Hugging Face platform to help developers improve documentation.

\subsection{Threats to Validity}  

\vspace*{0.2cm} 
\noindent \textbf{Threats to Internal Validity.} 
We use the keyword `\texttt{code}' to fetch an initial list of models from Hugging Face.
However, the returned results may not cover all LLM4Code.
To mitigate this threat, we follow the idea in the systematic literature review to conduct snowballing search to identify more code datasets and LLM4Code.
To confirm whether a dataset or model is relevant to code, we manually check their documentation.
However, as shown in our results, some datasets and models are not well documented, making it difficult to confirm their relevance to code.
We exclude datasets and models without clear documentation.

\vspace*{0.2cm}
\noindent \textbf{Threats to External Validity.} 
LLM4Code in this study are collected from Hugging Face.
However, the observation drawn from our study may not generalize to other LLM4Code on other model zoos.
Although this threat is inevitable, we believe that our study can still provide valuable insights for the community considering the popularity of Hugging Face and the dominant position of Transformers-based models in processing software engineering tasks.
Besides, the field of deep learning models for software engineering is evolving rapidly. 
New patterns in the ecosystem can emerge, with more models be published and new reuse patterns be developed.
We show our findings can generalize to the expanded dataset and plan to extend and revisit the ecosystem periodically.

\section{Related Work} \label{sec:related}

We refer interested readers to a recent survey~\cite{codellm_survey} for a comprehensive review of LLM4Code. 
In this section, we explain the related works on analyzing AI model zoos to understand AI development practices.

With the increasing popularity of `pre-training' and `fine-tuning' paradigm in deep learning, researchers have started to explore the reuse in deep learning models.
Jiang et al.~\cite{jiang2023empirical} analyze the models hosted on Hugging Face and identify useful attributes (e.g, reproducibility and portability) and challenges (e.g., missing attributes and discrepancies) for model reuses.
Taraghi et al.~\cite{10589852} analyze the challenges, benefit, and trends in reuse models on Hugging Face.
Davis et al.~\cite{davis2023empirical} divide the model reuse into high-level types: conceptual reuse, adaption reuse, and deployment reuse. 
In our paper, we categorize how LLM4Code are reused to derive new models at lower granularity, e.g., model conversion, distillation, adversarial training, etc.

Existing studies~\cite{10.1145/3643916.3644412,jones2024knowhuggingfacesystematic,10555709} highlights the lack of documentation Hugging Face models, which aligns with our observations on LLM4Code.
For example, Pepe et al.~\cite{10.1145/3643916.3644412} notice that the models are not transparent to document the models' training data.
Our study conducts analysis of multiple documentation practice, including the use of lists, code blocks, etc.
We also compare the documentation practice of LLM4Code with AI repositories hosted on GitHub~\cite{fan2021makes}.
Jiang et al.~\cite{PeaTMOSS} construct mappings between the Hugging Face models and GitHub repositories that use these models. 
They analyze the license usage of models and GitHub repositories to identify license incompatibility issues.
Different from Jiang et al.~\cite{PeaTMOSS}, we build dependencies between models with reuse relationships and see whether the license of the derived model is compatible with the base model.

Researchers have also explored other aspects of models on Hugging Face, including carbon footprint~\cite{10304801}, bias~\cite{10.1145/3643916.3644412}, analyzing methodology~\cite{10.1145/3643664.3648204}, data privacy and security~\cite{wang2024largelanguagemodelsupply}, naming convention~\cite{jiang2024namingpracticespretrainedmodels}, etc.
Specifically, researchers~\cite{jiang2022empirical,davis2023empirical,10297271} have also noticed the potential security risks of reusing deep learning models, e.g., vulnerabilities to backdoor attacks.
In the context of LLM4Code, researchers~\cite{codebackdoor,advdoor,you-see} demonstrate that attackers can easily poison training data and inject backdoors to LLM4Code to manipulate their behaviors.
It further highlights the importance of building the dependencies between models and datasets to track the training data and identify potential security risks.

There is one study focusing on LLM4Code hosted on Hugging Face.
Sipio et al.~\cite{10.1145/3661167.3661215} build a taxonomy of software engineering tasks and develop an approach to categorize the pre-trained models for software engineering.
Our work differs from them by focusing on the dependency, documentation, and licensing practices in the LLM4Code ecosystem.
The taxonomy developed in Sipio et al.'s study~\cite{10.1145/3661167.3661215} may benefit the ecosystem analysis as well, e.g., understand the tasks distribution and the corresponding trends in the LLM4Code ecosystem.
We leave it as future works.

\section{Conclusion and Future Work} \label{sec:conclusion}

This study presents the first comprehensive exploration of the LLM4Code ecosystem. 
By manually curating a dataset of 366 models and 73 datasets from Hugging Face, we gain significant insights into the ecosystem's structure and dynamics. 
Our analysis shows the disparity between company and non-company contributions.
We develop a taxonomy of model reuse practices, highlighting strategies such as fine-tuning and architecture sharing, while identifying emerging approaches like quantization and adversarial tuning that address deployment and robustness challenges. 
Our examination of documentation and licensing practices also underscores the need for improved transparency and compatibility to support sustainable growth.
Furthermore, we demonstrate the potential of LLMs to automate the construction and analysis of the ecosystem. 
Based on our findings, we provide suggestions to stakeholders and identify future research directions to enhance the ecosystem's efficiency, transparency, and legal compliance.

This first exploration of the LLM4Code ecosystem opens opportunities for a series of future works. We plan to develop more accurate automated methods to construct and analyze the LLM4Code ecosystem at scale. It is also promising to consider more stakeholders and entities in the ecosystem, including the software that uses LLM4Code (e.g., IDEs with LLM4Code as plugins). 
We plan to mining the ecosystem to understand more critical questions like potential security risks in model reuse, enhancing model documentation, etc.

\bibliographystyle{ACM-Reference-Format}
\bibliography{./reference}

\end{document}